\documentclass[12pt,preprint]{aastex}
\usepackage{graphicx}
\begin{document}

\title{Angular Momentum Transfer in Vela-like Pulsar Glitches}

\author{Pierre M. Pizzochero}
\affil{Dipartimento di Fisica, Universit\`a degli Studi di Milano,
 and Istituto Nazionale  di Fisica Nucleare, sezione di Milano,  Via
 Celoria 16, 20133 Milano, Italy}
\email{pierre.pizzochero@mi.infn.it}

\begin{abstract}
The angular momentum transfer associated to Vela-like glitches has never been calculated {\em directly} within a realistic scenario for the storage and release of superfluid vorticity; therefore, the explanation of giant glitches in terms of vortices has not yet been tested against observations. We present the first physically reasonable model, both at the microscopic and macroscopic level (spherical geometry, n=1 polytropic density profile, density-dependent pinning forces compatible with vortex rigidity), to determine where in the star the vorticity is pinned, how much of it, and for how long. For standard neutron star parameters ($M=1.4 M_{\odot}, R_s=10$ km,    $\dot{\Omega}=\dot{\Omega}_{\rm Vela}=-10^{-10}$ Hz s$^{-1}$), we find that maximum pinning forces of order $f_m\approx10^{15}$ dyn cm$^{-1}$ can accumulate  $\Delta L_{\rm gl}\approx10^{40}$ erg s of superfluid angular momentum, and release it to the crust at intervals $\Delta t_{\rm gl}\approx3$ years. This estimate of $\Delta L_{\rm gl}$ is  one order of magnitude smaller than what implied indirectly by current models for post-glitch recovery,  where the core and inner-crust vortices are taken as physically disconnected; yet, it successfully yields the magnitudes observed in recent Vela glitches for {\em both} jump parameters, $\Delta\Omega_{\rm gl}$ and $\Delta\dot{\Omega}_{\rm gl}$, provided one assumes that only a small fraction ($<10\%$) of the total star vorticity  is coupled to the crust on the short timescale of a glitch. This is reasonable in our approach, where no layer of normal matter exists between the core and the inner-crust, as indicated by existing microscopic calculation.  The new scenario presented here is nonetheless compatible with current post-glitch models.
\end{abstract}

\section{Introduction}

Glitches are sudden spin-ups observed in the otherwise decreasing rotational frequency of a pulsar \citep{lyn00}. Their origin is still debated: the giant spin-ups observed in the twenty known Vela-like glitchers \citep{esp11} could indicate the presence of bulk superfluidity inside these stars. In this scenario, giant glitches would represent the natural macroscopic outcome of the interaction between quantized neutron vortex lines, which carry the angular momentum of the rotating chargeless superfluid, and the Coulomb lattice of neutron-rich nuclear clusters, which coexists with the neutron superfluid in the inner crust \citep{nv73}. Indeed, this interaction can pin vortices to the normal component of the star, thus freezing the superfluid vorticity and storing its angular momentum. Only when the hydrodynamical lift on a vortex (Magnus force), which increases as the pulsar slows down, equals the pinning force on a line, the vortex is unbound from the lattice: thus free to move under the action of drag forces, it can transfer its angular momentum to the normal component of the star. According to \citet{ai75}, giant glitches are due to  the sudden and simultaneous depinning of a large number of accumulated vortices, followed by the rapid transfer of their angular momentum to the observable {\em normal} crust (which consists of the outer crust plus all the other charged components in the star, electrons, protons, and nuclear clusters, strongly coupled together by the pulsar magnetic field). Such a storage and trigger mechanism would have a natural periodicity, as indeed observed in Vela \citep{dod07}.

The vortex scenario for glitches was roughly compared to existing observations in a simple but instructive toy-model, which assumed a cylindrical, uniform-density star with a cylindrical pinning shell (corresponding to the inner-crust) close to the surface \citep{pin80, alp81, and82}. Although the naive treatment of the vortex-nucleus and vortex-lattice interactions gave pinning forces three orders of magnitude larger than what required to explain the average interval between glitches observed in Vela ($\Delta t_{\rm gl}\approx3$ years),  the model predicted the correct orders of magnitude for the typical glitch parameters known at the time, namely  the jump in angular velocity, $\Delta\Omega_{\rm gl}\approx10^{-6}\Omega$, and the jump in angular acceleration, $\Delta\dot{\Omega}_{\rm gl}\approx10^{-2}\dot{\Omega}$  (the pre-glitch, steady-state parameters for Vela being $\Omega_{\rm Vela}=70$ Hz and $\dot{\Omega}_{\rm Vela}=-9.8\times10^{-11}$ Hz s$^{-1}$). In spite of these positive preliminary results, \citet{pin80} carefully pointed out that the effects of some crucial corrections had to be taken into account before drawing any conclusion: the spherical geometry of the star, the radial density profile required by gravitational equilibrium, the density dependence of the pinning interaction, and the presence of different superfluid phases along the star profile. To date, however, this has not been done in any coherent and consistent model; thus, the explanation of giant glitches in terms of vortices is not yet tested against observations, leaving the origin of these spin-ups a still open question.

The post-glitch recovery of pulsars, on the other hand, has been successfully interpreted in terms of vortex motion under drag forces. Early on, the phenomenological model of \citet{bay69}  explained the slow relaxation to steady-state following a glitch as due to the weak interaction between a normal and a superfluid component, each rotating rigidly. Following a glitch, the response of the model is linear in the initial perturbation, and relaxes back to steady-state conditions exponentially,  with a  relaxation time which is inversely proportional to the strength of the interaction between the two components. The simple two-components model was then reformulated in terms of vortex motion, to allow for differential rotation of the superfluid  \citep{pin80}.  Eventually, two scenarios were developed to describe vortex dynamics between glitches: thermally-activated creep of strongly pinned vortices \citep{alp84a,alp84b,lin93} or corotation of unpinned vortices under weak drag forces \citep{jon90, jon91,jon93}.  The vortex creep model was motivated by the large pinning forces obtained in early calculations \citep{alp77,eb88}. Later on, however, the microscopic vortex-nucleus interaction was shown to be one order of magnitude smaller than what found earlier \citep{dpI,dpII,dpIII}. Moreover, \citet{jon92} argued that the mesoscopic vortex-lattice interaction, necessary to calculate the macroscopic pinning force on a vortex line, is likely to be a factor $\alpha_1\sim10^{-2}$ smaller than what naively assumed in early calculations, due to the random orientation of the macro-crystals forming the inner crust, as well as to the rigidity of vortex lines on distances of order $10^2-10^3R_{\rm ws}$ (with $R_{\rm ws}$ the radius of the Wigner-Seitz cells describing the nuclear lattice). Significantly smaller pinning forces favor the corotation model, where unpinned vortex lines  are weakly coupled to the normal crust by small drag forces; thence,  in steady-state they (nearly) corotate with the superfluid, while the response  to perturbations is linear.   Finally, the Christmas 1988 Vela glitch \citep{fl90} showed that the creep model can fit observations only in the linear response regime \citep{alp89}, in which case it is  equivalent to the corotation model, but with a different temperature-dependence of the drag parameters. Observationally, the post-glitch recovery of Vela is well described by a sum of exponential terms with different amplitudes and relaxation times \citep{dod02}, and this can be explained in terms of linear response of regions of the superfluid characterized by different drag parameters. The dissipative force has been evaluated for several densities of interest, and the corresponding drag parameters yield relaxation times and glitch rise-times compatible with observations \citep{jon90,jon92,eb92}.
   
Three aspects of current post-glitch models  are relevant here: \\
{\em i)}  although simulations successfully reproduce the observed recovery of Vela \citep{lar02}, the glitch itself is always introduced by hand, as an ad hoc initial condition. \\
{\em ii)}  core and crust vortices are taken as physically disconnected, namely a layer of normal matter is {\em assumed} between the S-wave neutron superfluid found at subnuclear densities and the P-wave one found above nuclear saturation \citep{jon90}.  Recent microscopic calculations, however, do not show any discontinuity of this kind \citep{zgap04}, thus indicating continuous vortex lines throughout the star. \\
{\em iii)}  in the core, the superconducting state of protons determines in part the mutual friction on neutron vortex lines. A type II superconductor corresponds to the strong-drag limit, with vortices entangled in a dense array of magnetic flux tubes \citep{lin03}; for a type I superconductor, instead, both the weak- and strong-drag limits have been suggested \citep{sed05,jon06}. To date, the microscopic nature of the proton superconductor is far from settled; therefore, both scenarios of weak and strong mutual friction in the core should be taken into account in the study of glitches. Post-glitch models, however,  are not affected by this theoretical uncertainty, since they involve mostly vortices in the equatorial regions, lying entirely in the inner crust.

A typical neutron star has total momentum of inertia  $I_{\rm tot}\approx10^{45}$ g cm$^2$, while its inner crust has $I_{\rm ic}\approx10^{-2}I_{\rm tot}$. The accidental coincidence of the ratio $I_{\rm ic}/I_{\rm tot}$ with the early observations of $\Delta\dot{\Omega}_{\rm gl}\approx10^{-2}\dot{\Omega}$ and with fact that about $1.7\%$ of the Vela spin-down is reversed during a glitch, led to consider glitches as related to crust vorticity alone, and thence to the assumption of disconnected neutron superfluids.  This scenario, however, has direct implications on the glitch energetics. Indeed, since vortex lines in the core are strongly coupled to the normal component, being magnetized by entrainment effects \citep{alp84c}, the normal crust comprises most of the star  and $I_{\rm c}\approx I_{\rm tot}$. This implies that the angular momentum transferred during the glitch is $\Delta L_{\rm gl}=I_{\rm c}\Delta\Omega_{\rm gl}\approx10^{41}$ erg s, and the corresponding glitch energy is $\Delta E_{\rm gl}=\Delta L_{\rm gl}\Omega\approx10^{43}$ erg; both values appear too large. On the one hand,  $10^{41}$ erg s  corresponds to the difference in angular momentum of the entire inner crust between glitches, thus requiring some unlikely mechanism that freezes the vorticity {\em everywhere} in the crust for about 3 years, and then releases it simultaneously. On the other hand, observations of the power wind nebula surrounding Vela indicate an upper limit of  $\sim10^{42}$ erg to the glitch energy  \citep{hel01}.  

In this letter we present  the first realistic (with several approximations, but still preserving the essential physics) and consistent model to determine where in the star the vorticity is pinned, how much of it, and for how long. The model has been tested against observations using realistic equations of state (EoS) for dense matter and implementing general relativistic hydrostatic equilibrium (Pizzochero, Seveso  \& Haskell, in preparation). Moreover, initial dynamical simulations based on the multifluid formalism of \citet{ac06} confirm the main assumptions and predictions of the model \citep{hps11}.  Here, however, we will discuss a fully analytical, Newtonian version of the model: it yields the correct orders of magnitude for all relevant variables, but provides deeper insight than any numerical treatment.

\section{The model}

 We now outline the main assumptions of the model and present the resulting equations; details and calculations, together with a parameter study of the solutions, will be given in a longer article (Pizzochero, in preparation; from now on, Paper I).  
 
\begin{figure}
\includegraphics{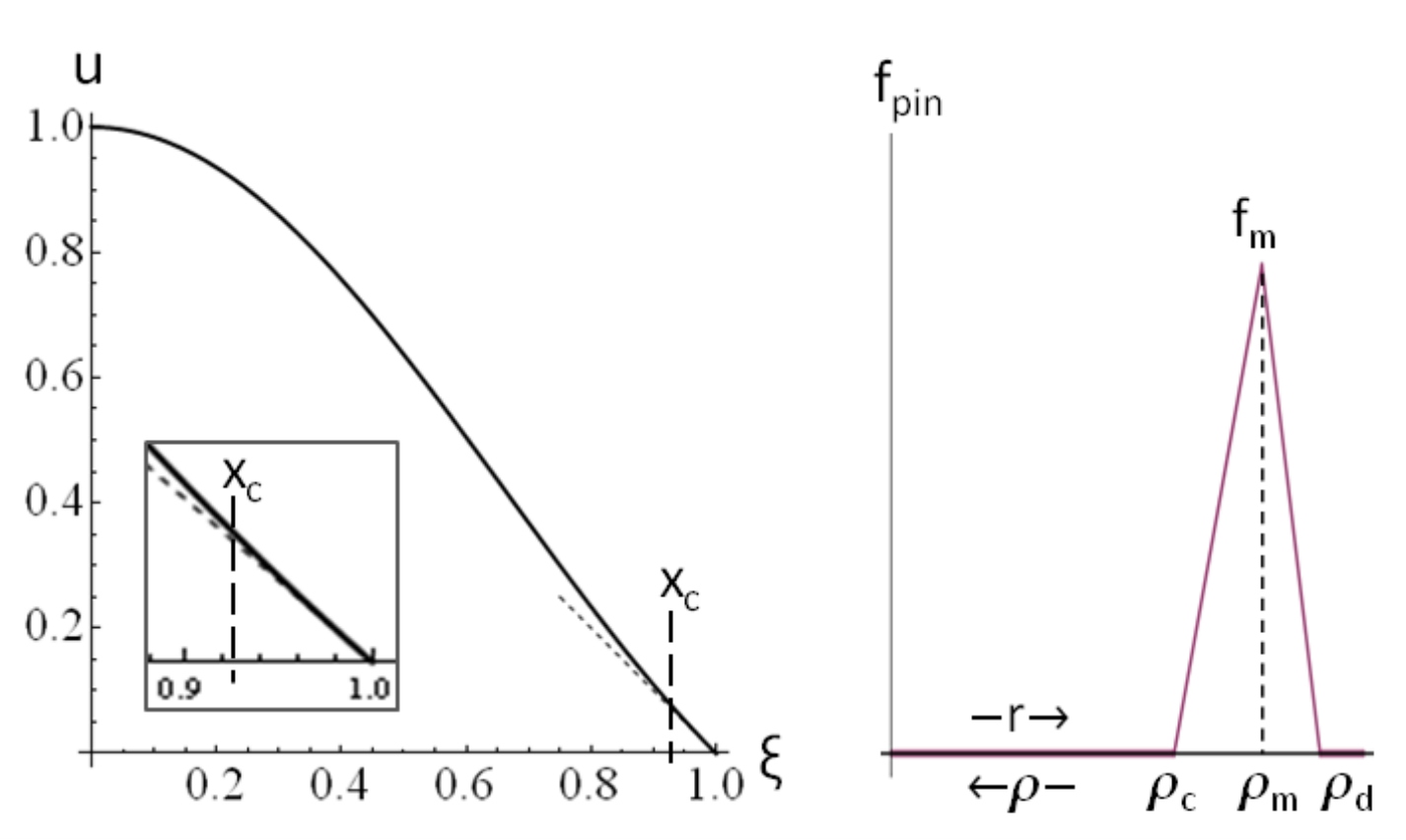}
\caption{(left) Density profile for the n=1 polytrope in dimensionless variables;  the dotted line is the low-density approximation $u=1-\xi$. In the insert, the low-density region is magnified. (right) Density dependence of the pinning force. \label{fig1}}
\end{figure}

 We describe the core and inner crust as a n=1 polytrope, of mass $M$ and radius $R_s$. The actual radius of the star will be larger, because of the overlying outer crust; its presence, however, can be ignored here, since it contributes negligibly to the mass and moment of inertia of the normal component. The polytropic relation $P\propto \rho^2$ is a very soft EoS for dense matter; realistic soft EoSs yield $R_s\approx10$ km for $M=1.4 M_{\odot}$. The density profile is $u=\sin(\pi \xi)/(\pi \xi)$, with the dimensionless radius $\xi=r/R_s$ and the  density $u=\rho/\lambda$, normalized to its central value $\lambda=\pi M/(4R_s^3)$.  The radius of the core, $R_c$, corresponds to the density $\rho_c=0.6\rho_0$ (in units of nuclear saturation density, $\rho_0=2.8\times10^{14}$ g cm$^{-3}$), where nuclei merge into nuclear matter. The inner crust has $\xi>x_c$, where $x_c=R_c/R_s>0.9$; for $\xi>0.9$, the approximation $u=1-\xi$ is sufficiently accurate (Figure \ref{fig1}, left). The total momentum of inertia is
 \begin{equation}
I_{\rm tot}=\frac{2(\pi^2-6)}{3\pi^2}MR_s^2=0.26MR_s^2, \label{eq1}
\end{equation}
while  the inner crust has  $M_{\rm ic}=4.9(1-x_c)^2M$ and $I_{\rm ic}=12.6(1-x_c)^2I_{\rm tot}$. 

The standard {\em superfluid fraction}, $Q$, is introduced to describe the protons and nuclei of the normal crust, $I_c=(1-Q)I_{\rm tot}$, and the neutron superfluid component, $I_s=QI_{\rm tot}$, of the star. Although the neutron fraction varies with density, a typical average value is $Q\approx0.95$ \citep{zxp04}. Regarding proton superconductivity in the core, here we choose the weak-drag limit. The model, however, gives reasonable results for the jump parameters also in the strong-drag limit; the decoupling of the core vorticity, pinned by flux-tubes, reduces both $\Delta L_{\rm gl}$ and the moment of inertia responding at the glitch (cf. Paper I).

The density dependence  of  $f_{\rm pin}(\rho)$, the (mesoscopic) pinning force per unit length, is taken as in Figure \ref{fig1}(right).  This is a reasonable first approximation for a parameter study in terms of the maximum value, $f_m$; indeed, pinning goes to zero at $\rho_c$ (no more nuclei) and at neutron drip $\rho_d=0.0015\rho_0$ (no more neutrons), while it is expected to be maximum around densities where the pairing gap peaks. Moreover, we have performed a numerical simulation to evaluate $f_{\rm pin}(\rho)$ in a bcc lattice, with random crystal orientations and proper vortex rigidity. We obtain profiles compatible with Figure \ref{fig1}, with maximum values of order $f_m\approx10^{15}$ dyn cm$^{-1}$ at densities $\rho_m\approx0.2\rho_0$. We also find, as already noted by \citet{lin09}, that attractive and repulsive vortex-nucleus interactions are equivalent for pinning vortices to the lattice (Grill \& Pizzochero, in preparation).

\begin{figure}
\includegraphics{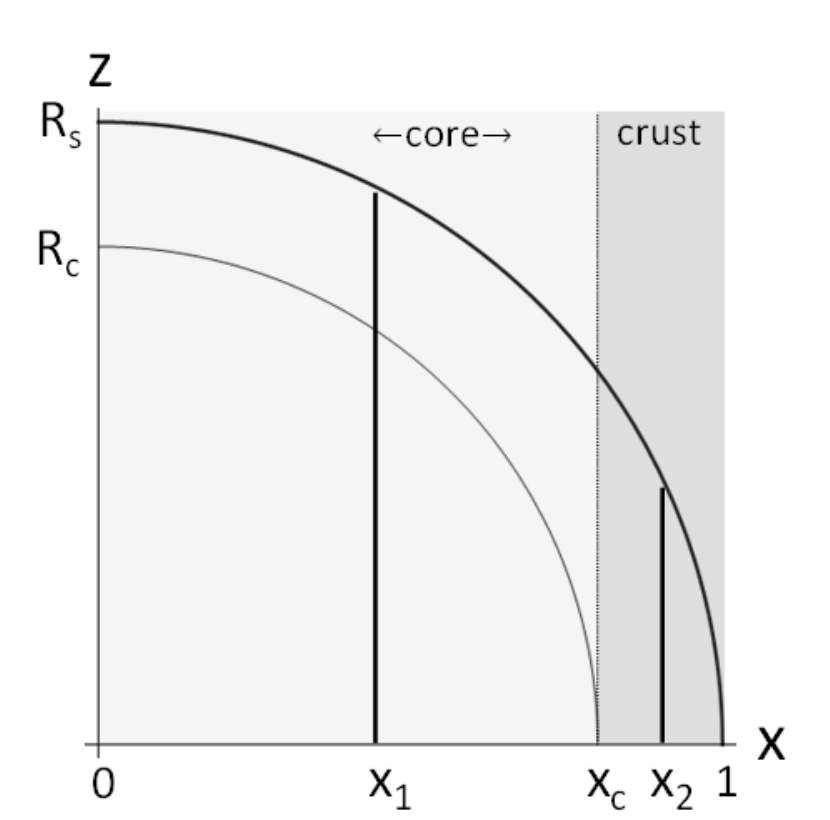}
\caption{Geometry of the model (not in scale): the star rotates around the $z$-axis, while the $x$-axis represents the  dimensionless radius in cylindrical geometry. Two vortices are shown at $x_1$ and $x_2$,  respectively in the \lq{core}\rq{} and the \lq{crust}\rq{}, the two cylindrical regions separated  by $x_c=R_c/R_s$ .
\label{fig2}}
\end{figure}

The geometry of the model is shown in Figure \ref{fig2}.  The star spins around the $z$-axis, and continuous vortex lines are assumed through the core. Following the results of \citet{rud74}, we can reduce the problem to axial symmetry by integrating the density-dependent quantities along the vortex lines; these quantities will then depend only on the {\em cylindrical} radius $x=R/R_s$, with $R$ the distance from the rotational axis.
We distinguish two cylindrical zones, separated  by $x_c=R_c/R_s$: the \lq{crust}\rq{} ($x>x_c$), with vortices lying entirely in the inner crust, and the  \lq{core}\rq{} ($x<x_c$), with vortices crossing the star core. In particular, we can integrate the pinning and Magnus forces to obtain an estimate of their total values on a vortex. If $\omega(x)=\Omega_s(x)-\Omega$ indicates the lag between the local superfluid angular velocity and that of the rigid normal crust, the critical lag for depinning, $\omega_{\rm cr}(x)$, is obtained by equating  these two forces
\begin{eqnarray}
\int_v{\rm d}z\,f_{\rm pin}[\rho(x,z)]=x\omega_{\rm cr}(x)\kappa\int_v{\rm d}z\,\rho(x,z),                                    
\end{eqnarray}
where $\kappa=\pi\hbar/m_N$.  

\begin{figure}
\includegraphics{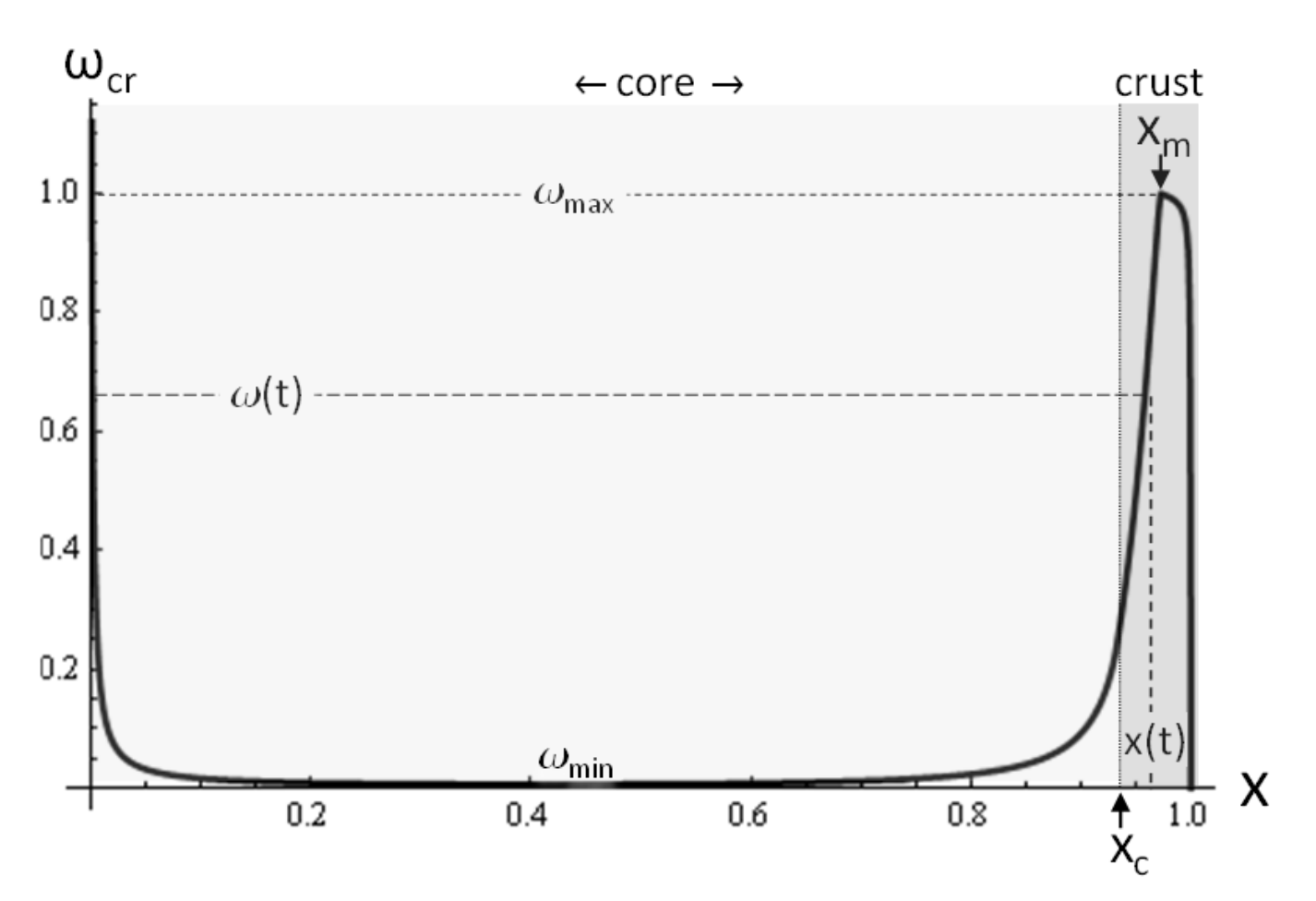}
\caption{Critical lag for depinning, $\omega_{\rm cr}(x)$, as a function of the dimensionless cylindrical radius $x$.  At any given time, the lag $\omega(t)$ determines the position  $x(t)$.  The lags have been normalized to $\omega_{\rm max}$.
 \label{fig3}}
\end{figure}

In Figure \ref{fig3}, we show the resulting profile: $\omega_{\rm cr}(x)$ presents a sharp peak in the \lq{crust}\rq{}, with maximum $\omega_{\rm max}$ located very close to $x_m=1-u_m=1-\rho_m/\lambda$; we will take $\rho_m=0.2\rho_0$. In most of the \lq{core}\rq{}, instead, $\omega_{\rm cr}(x)$ has a roughly uniform value $\omega_{\rm min}\approx10^{-2}\omega_{\rm max}$ (as $x\rightarrow0$ it diverges; similarly to the outer crust, however, this region can be neglected). We find (cf. Paper I)
\begin{eqnarray}
\omega_{\rm max}=\omega_{\rm cr}(x_m)=\frac{4}{\kappa}\frac{R_s^2}{M}\frac{g_{\rm pin}(x_m)}{g_{\rm mag}(x_m)}f_m,  \label{eq2}
\end{eqnarray}
where
\begin{mathletters}
\begin{eqnarray}
g_{\rm pin}(x)&=&\frac{1}{2(1-x)}\left[\sqrt{1-x^2}-x^2\ln\left(\frac{1+\sqrt{1-x^2}}{x}\right)\right] \\
g_{\rm mag}(x)&=&\pi x\left[\ln\left(\frac{1+\sqrt{1-x^2}}{x}\right)-\sqrt{1-x^2}\right].
\end{eqnarray}
\end{mathletters}

In the  \lq{crust}\rq{}, this estimate of $\omega_{\rm cr}(x)$ should be reasonable, since pinning is continuous along the vortices \citep{rud74,jon90}. In the  \lq{core}\rq{}, instead, pinning is discontinuous:  vortex lines are attached to the lattice only at their extremities, while most of their length lies in a pinning-free region (having selected the weak-drag limit). We can expect individual string-like excitations of the pinned vortices, which could detach them from the lattice well before  $\omega_{\rm min}$  is reached.  Indeed, the collective rigidity of vortex bundles in coherent motion, which explains the axial symmetry predicted by the Taylor-Proudman theorem \citep{rud74}, is actually not observed in laboratory experiments with superfluid vortices attached to the rotating vessel only at their ends \citep{ada85}. 

Although this issue requires and deserves further study, the crucial point is that vortices  in the  \lq{core}\rq{} are pinned very weakly. On the other hand, drag forces due to magnetization correspond to very short relaxation times and thence very small steady-state lags \citep{alp84c}. Moreover,  \citet{lin09} has shown that vortex repinning is dynamically possible if the lag falls below a critical value (smaller than the critical lag for depinning).  These considerations and the profile in Figure \ref{fig3} naturally suggest the following scenario: as the star slows down, vortices in the  \lq{core}\rq{} are continuously depinned and then rapidly repinned; this  {\em dynamical creep} allows a steady removal of the excess vorticity on short effective timescales $\tau_c$. Although the value of  $\tau_c$  is not relevant here, dynamical simulations of Vela glitches suggest  $\tau_c\sim10^0-10^1$ s  \citep{hps11}, compatible with mutual friction dominated by vortex magnetization. On timescales $\Delta t\lesssim\tau_c$  (e.g., during a glitch), the \lq{core}\rq{} vorticity is only partially coupled to the normal crust (the rest being pinned or responding on longer timescales), and only a detailed study of the dynamics can provide a direct estimate of the coupled fraction. On longer timescales $\Delta t\gg\tau_c$, however,  the dynamical creep ensures full effective coupling of the two components with (average) lag of order $|\dot{\Omega}|\tau_c$; in steady-state, this scenario is then equivalent to the corotation model.  

The  excess \lq{core}\rq{} vorticity will be repinned  in the \lq{crust}\rq{}, where pinning increases rapidly by orders of magnitude.  At any time $t$ after a glitch, the lag $\omega(t)=|\dot{\Omega}_{\infty}|t$ defines a radial distance $x(t)$ as in Figure \ref{fig3}; here $\dot{\Omega}_{\infty}$ indicates the steady-state (pre-glitch) angular acceleration. We now assume that the excess vorticity, corresponding to the entire region $x<x(t)$ and to the lag $\omega(t)$, is accumulated in a thin vortex sheet at $x(t)$; as the star slows down and $\omega(t)$ increases, the sheet is pushed outwards by the increasing Magnus force, and moves with $x(t)$. When $\omega(t)$ reaches the value $\omega_{\rm max}$, the sheet is at the pinning peak, $x_m$, and the vorticity accumulated in the sheet is finally released simultaneously, causing the glitch. This picture reminds of a snowplow, pushing accumulated snow up an incline and eventually reaching its top edge. The interval between glitches is 
\begin{eqnarray}
\Delta t_{\rm gl}=\frac{\omega_{\rm max}}{|\dot{\Omega}_{\infty}|}. \label{eq4}
\end{eqnarray}  
 This scenario is compatible with post-glitch relaxation; indeed, after a glitch, the unpinned vortices in the \lq{crust}\rq{} are under the same conditions as those considered in current post-glitch models. 

\begin{figure}
\includegraphics{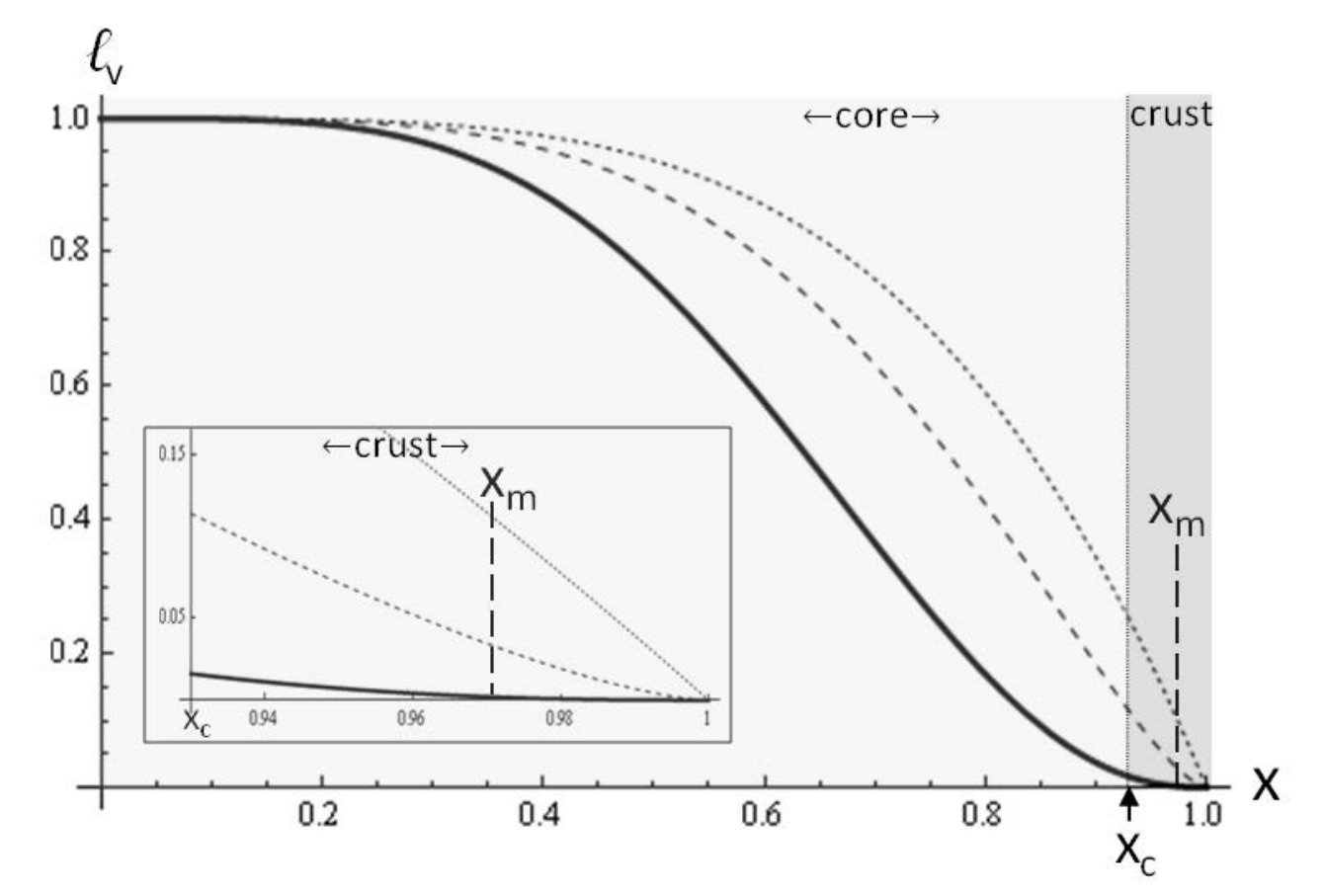}
\caption{Reduction of angular momentum, $\ell_v(x)=L_v(x)/L_v(0)$,  when uniformly distributed vorticity contained within $x$ is accumulated into a vortex sheet at $x$. Three scenarios are considered: uniform-density cylindrical star (dotted), uniform-density spherical star (dashed), n=1 polytrope (solid). In the insert, the \lq{}crust\rq{} region is magnified.   \label{fig4}}
\end{figure}

Although the \lq{snowplow}\rq{} model is quite schematic, it contains a plausible mechanism for storing and releasing vorticity, as actually confirmed by parallel dynamical simulations \citep{hps11}. In particular, the model allows to calculate {\em directly}  the angular momentum $L_v(x)$ of the vortex sheet at $x$. In Figure \ref{fig4} we show the reduction of angular momentum, $\ell_v(x)=L_v(x)/L_v(0)$, when uniformly distributed vorticity contained within $x$ is accumulated in a sheet at $x$; at the peak, $x_m$, the reduction is of order $10^{-3}$. For comparison, we also show the significantly different results for a uniform-density, cylindrical or spherical star; we  see how spherical symmetry and realistic density profile are {\em both} crucial to obtain the correct order of magnitude of $\ell_v(x)$. 

The angular momentum stored during $\Delta t_{\rm gl}$ and released at the glitch, $\Delta L_{\rm gl}$,  can be calculated from the number of vortices removed from the interior and accumulated at $x_m$, namely $\Delta N_v(x_m)=2\pi R_s^2x_m^2\omega_{\rm max}/\kappa$.  We find (cf. Paper I)
\begin{eqnarray}
\Delta L_{\rm gl}=I_{v}(x_m)\omega_{\rm max},  \label{eq5}
\end{eqnarray}
with an effective moment of inertia 
\begin{eqnarray}
I_{v}(x)=\frac{3\pi^4}{2(\pi^2-6)}g_{v}(x)QI_{\rm tot}=\pi^2g_{v}(x)QMR_s^2, \label{eq6}
\end{eqnarray}
where
\begin{eqnarray}
g_{v}(x)=\frac{x^2}{6}\left[\sqrt{1-x^2}\left(1+2x^2\right)-3x^2\ln\left(\frac{1+\sqrt{1-x^2}}{x}\right)\right]. \label{eq7}
\end{eqnarray}

The glitch rise-time is very short, $\tau_{\rm gl}<40$ s \citep{dod02}; we introduce a new parameter, $Y_{\rm gl}$, which globally describes the {\em fraction} of vorticity {\em coupled} to the normal crust on timescales of order $\tau_{\rm gl}$ (the steady-state coupled fraction, corresponding to long timescales and to pre-glitch conditions,  is $Y_{\infty}=1$). The value of $Y_{\rm gl}$ depends on the detailed short-time dynamics of the \lq{core}\rq{} vorticity; in order to get an estimate of the observables, only this quantity is needed. From angular momentum conservation and variation of the crust equation of motion we find the glitch jump parameters 
\begin{mathletters}
\begin{eqnarray}
\Delta\Omega_{\rm gl}&=&\frac{\Delta L_{\rm gl}}{I_{\rm tot}[1-Q(1-Y_{\rm gl})]}  \\
\frac{\Delta\dot{\Omega}_{\rm gl}}{\dot{\Omega}_{\infty}}&=&\frac{Q(1-Y_{\rm gl})}{1-Q(1-Y_{\rm gl})}. \label{eq9}
\end{eqnarray}
\end{mathletters}

\section{Results and observations}

After fixing the basic stellar parameters $M$, $R_s$ and $Q$ (more generally, $M$ and an EoS), the model has two free parameters, $f_m$ and $Y_{\rm gl}$. It must predicts three observables: the interval between glitches, and the jumps in angular velocity and acceleration during a glitch. In the case of Vela, the average observed values are $\Delta t_{\rm gl}\approx3$ years and $\Delta\Omega_{\rm gl}=1.2\times10^{-4}$ Hz \citep{lyn00}; we already mentioned that early observations gave $\Delta\dot{\Omega}_{\rm gl}/\dot{\Omega}_{\infty}\approx10^{-2}$.  More recent data, however, indicate much larger values; in particular,  the year 2000 glitch \citep{dod02} added to the  already known  short-, middle-, and long-time relaxation components ($\tau_i\approx10^4,10^5,10^6$ s, with $\Delta\dot{\Omega}_i/\dot{\Omega}_{\infty}\approx0.44,0.044,0.009$ for $i=1,2,3$),  a fourth  and {\em very} short one, with $\tau_4=1.2\pm0.2$ minutes and $\Delta\dot{\Omega}_4/\dot{\Omega}_{\infty}=18\pm6$ (one sigma errors). In the 2004 glitch, however, such a component was observed only barely above noise  and no firm conclusion could be drawn from the weak data \citep{dod07}. Waiting for future observations, there is nonetheless evidence that {\em right after} a glitch $\Delta\dot{\Omega}_{\rm gl}/\dot{\Omega}_{\infty}$  is larger than unity. 

In order to test the model against observations, we consider a standard neutron star with $M=1.4 M_{\odot}, R_s=10$ km, and $Q=0.95$. If we take $f_m=1.1\times10^{15}$ dyn cm$^{-1}$, from equations \ref{eq2}$-$\ref{eq7} we find  that $\omega_{\rm max}=0.01$ Hz, and thence $\Delta t_{\rm gl}=3.1$ years and $\Delta L_{\rm gl}=9.5\times10^{39}$ erg s (also, $\Delta E_{\rm gl}=6.7\times10^{41}$ erg).  If we then take $Y_{\rm gl}=0.05$, from equation 9 we obtain $\Delta\Omega_{\rm gl}=1.3\times10^{-4}$ Hz and $\Delta\dot{\Omega}_{\rm gl}/\dot{\Omega}_{\infty}=9.3$, in good general agreement with observations. In Paper I we analyze the parameter dependence of these results; we find that the model is quite robust under physically meaningful variations of all the basic parameters ($M,R_s,Q,\rho_c,\rho_m,\rho_d$).

In conclusion,  assuming continuous vortices throughout the star, we find that maximum pinning forces of order $f_m\approx10^{15}$ dyn cm$^{-1}$ can accumulate $\approx10^{13}$ vortices in the inner crust of a standard neutron star, and release them every $\approx3$ years, transferring an angular momentum  $\Delta L_{\rm gl}\approx10^{40}$ erg s. This is one order of magnitude smaller than what inferred from the (microscopically inconsistent) assumption of disconnected vortices. Yet, it yields the observed glitch parameters, provided one assumes a small coupled fraction $Y_{\rm gl}<10\%$. The model is compatible with post-glitch recovery and with the presently known microphysics; the numerical results follow from implementing both spherical geometry and a realistic density profile, and they are robust.

\acknowledgments
 
This work was supported by CompStar, a Research Networking Programme of the European Science Foundation (\url{http://www.compstar-esf.org/}).

\clearpage

\clearpage

\clearpage

\clearpage
\end{document}